\documentclass[reprint, superscriptaddress, amsmath, amssymb, aps, longbibliography, nofootinbib]{revtex4-1}

\usepackage[english]{babel}
\usepackage[utf8x]{inputenc}
\usepackage[T1]{fontenc}

\usepackage{mathtools}
\DeclarePairedDelimiter\abs{\lvert}{\rvert}
\DeclarePairedDelimiter\norm{\lVert}{\rVert}
\usepackage{graphicx}
\usepackage{subfig}
\usepackage{amsmath}
\usepackage{physics}
\usepackage{outlines}
\usepackage[colorinlistoftodos]{todonotes}
\usepackage[colorlinks=true, allcolors=blue]{hyperref}
\usepackage{booktabs}

\usepackage{color,soul}

\newcommand{\kket}[1]{| #1 \rangle}

\DeclarePairedDelimiterX{\infdivx}[2]{(}{)}{%
  #1\;\delimsize\|\;#2%
}
\newcommand{\kldiv}{D_{\text{KL}}\infdivx}

\begin{document}

%% Title page/section
\title{Expressibility and entangling capability of parameterized quantum circuits for hybrid quantum-classical algorithms}

\author{Sukin Sim}
\email{ssim@g.harvard.edu}
\affiliation{%
 Department of Chemistry and Chemical Biology, Harvard University, 12 Oxford Street, \protect\\ Cambridge, MA 02138, USA
}%
\affiliation{%
 Zapata Computing, Inc., 501 Massachusetts Avenue, \protect\\ Cambridge, MA 02139, USA
}%

\author{Peter D. Johnson}
\affiliation{%
 Zapata Computing, Inc., 501 Massachusetts Avenue, \protect\\ Cambridge, MA 02139, USA
}%

\author{Al\'{a}n Aspuru-Guzik}%
\email{alan@aspuru.com}
\affiliation{%
 Zapata Computing, Inc., 501 Massachusetts Avenue, \protect\\ Cambridge, MA 02139, USA
}%
\affiliation{%
 Department of Chemistry and Department of Computer Science, University of Toronto, 80 St. George Street, \protect\\ Toronto, ON M5S 3H6, Canada
}%
\affiliation{%
 Canadian Institute for Advanced Research (CIFAR) Senior Fellow, 661 University Avenue, Suite 505, \protect\\ Toronto, ON M5G 1M1, Canada
}%
\affiliation{%
 Vector Institute, 661 University Avenue, Suite 710 \protect\\ Toronto, ON M5G 1M1, Canada
}%

\date{\today}

\begin{abstract}
Parameterized quantum circuits play an essential role in the performance of many variational hybrid quantum-classical (HQC) algorithms.
One challenge in implementing such algorithms is to choose an effective circuit that well represents the solution space while maintaining a low circuit depth and number of parameters.
To characterize and identify expressible, yet compact, parameterized circuits, we propose several descriptors, including measures of expressibility and entangling capability, that can be statistically estimated from classical simulations of parameterized quantum circuits.
We compute these descriptors for different circuit structures, varying the qubit connectivity and selection of gates. From our simulations, we identify circuit fragments
that perform well with respect to the
descriptors.
In particular, we quantify the substantial improvement in performance of two-qubit gates in a ring or all-to-all connected arrangement compared to that of those on a line.
Furthermore, we quantify the improvement in expressibility and entangling capability achieved by sequences of
controlled X-rotation
gates compared to sequences of 
controlled Z-rotation
gates.
In addition, we investigate how expressibility ``saturates'' with increased circuit depth, finding that the rate and saturated-value appear to be distinguishing features of a parameterized quantum circuit template.
While the correlation between each descriptor and performance of an algorithm remains to be investigated, methods and results from this study can be useful for both algorithm development and design of experiments for general variational HQC algorithms.
\end{abstract}
\maketitle

Due to significant development in both algorithm and hardware, we are expected to approach the era of ``Noisy Intermediate-Scale Quantum'' (NISQ) devices in the near future \cite{Preskill2018}. This generation of quantum machines are expected to support $50-100$ qubits and around $10^3$ gate operations. While these devices cannot perform error-corrected, large scale quantum computations, smaller but meaningful computations are anticipated to find use by combining both quantum and classical computational resources.

A particular class of algorithms that maximizes the use of such pre-threshold hardware is the hybrid quantum-classical (HQC) algorithm, which strategically divides computational tasks between quantum and classical resources. A prime example of a HQC algorithm is the variational quantum eigensolver (VQE), used to compute the ground states of molecular systems \cite{McClean2016, Cao2018}. Within the VQE framework, a parameterized trial wavefunction of the system-of-interest is prepared on the quantum computer by tuning a quantum circuit based on the chosen parameterization. This is followed by measurements of the energy expectation value with respect to the ansatz parameters. These parameters are updated using optimization routines on the classical computer and fed back into the quantum device to prepare a ``better-learned'' trial state. This cycle of updating the trial state and its parameters continues until the convergence criteria (e.g. energy convergence) are satisfied.

Other examples of HQC algorithms include the quantum approximate optimization algorithm (QAOA) \cite{Farhi2014}, quantum autoencoder (QAE) \cite{Romero2017}, quantum variational error corrector (QVECTOR) \cite{Johnson2017}, classification via near-term quantum neural networks (QNN) \cite{Farhi2018, Havlicek2018, SchuldCircuitCentric}, generative modeling \cite{Dallaire-Demers2018, Lloyd2018, Zhu2018}, among others. 
These algorithms provide an approach for a variety of problems. Though their objective functions (e.g. energy, average fidelity, mean squared error, KL divergence) may differ, these algorithms share a quantum subroutine for producing parameterized trial state(s) or ansatz(e) such that the parameters can be tuned to correspond to the optimal or near-optimal objective function value. Consequently, the performance (i.e. accuracy, scalability) of each algorithm will depend on the expressive power of the chosen parameterized quantum circuit.

Despite the crucial role of the parameterized circuit for a variety of HQC algorithms, there is a lack of general understanding and intuition behind characteristics associated with an ``effective'' training circuit. For instance, given two circuits $A$ and $B$, which circuit is more suitable for a given application and why? Are there figures-of-merit, or more neutrally, descriptors we can compute for these circuits to approach such question? In this work, we present a set of operational descriptors for characterizing and evaluating parameterized quantum circuits using classical simulations such that given a particular problem to solve using a HQC algorithm, we are equipped with tools for making decisions about and designing parameterized quantum circuits.

Similar questions have been posed and addressed in the field of statistical learning, in which various metrics, such as empirical risk, generalization ability, and sample complexity,
have been proposed
to evaluate the performance of a learning algorithm
\cite{Shalev-Shwartz2014}.
In practice, a method used to improve performance defined by one metric can come at the cost of losing performance defined by another metric, demonstrating a tension between these descriptors.
One example is the balance between minimizing empirical risk and maximizing generalization ability for a supervised learning task: while the algorithm should learn the training data well, it is also important to perform well with respect to unobserved data.
In recent years, the theory has been extended to the quantum domain \cite{Arunachalam2017}.
Our work can be viewed as an effort to develop NISQ analogues to the
ansatz descriptors
in statistical learning theory.
Further investigation is needed to more rigorously connect descriptors from this work to analogous metrics in statistical learning theory.
But, 
situating our framework as a potential NISQ analogue %allows for 
urges us to investigate new questions about PQCs
and to address analogous challenges, such as the phenomenon of ``barren plateaus'' \cite{McClean2018} in training landscapes.

Before presenting the descriptors, we provide a brief background on the structure and past studies of parameterized quantum circuits in Section \ref{sec:parameterized_circuits}. In Section \ref{sec:circuit_descriptors}, we propose two descriptors, namely \emph{expressibility} and \emph{entangling capability}, that provide a quantitative description of parameterized circuits independent of the algorithm or application. In Section \ref{sec:numerical}, we demonstrate the utility of such descriptors by computing and analyzing these quantities for a select set of parameterized circuits, several of which have been designed or inspired by past studies. In Section \ref{sec:discussion}, we discuss key observations from our simulations, including 
the phenomenon of \emph{expressibility saturation},
the efficacy of two-qubit controlled Z-rotation ($\texttt{CRZ}$) versus controlled X-rotation ($\texttt{CRX}$) gates, and the circuit configuration or topology. We conclude with a summary and outlook of future directions for this work.

\section{\label{sec:parameterized_circuits}Parameterized quantum circuits}
We define a parameterized quantum circuit (PQC) as a tunable unitary operation on $n$ qubits, $ U_{\boldsymbol\theta}$, that is applied to a reference state $\ket{\phi_{0}}$, often set to $\ket{0}^{\otimes n}$. The resulting parameterized quantum state is:

\begin{align} \label{eq:avgfid}
\ket{\psi_{\boldsymbol\theta}} = U_{\boldsymbol\theta} \ket{\phi_{0}},
\end{align}

\noindent where $\boldsymbol\theta$ is a vector of a polynomial number of circuit parameters. In this work, circuit parameters correspond to angles of rotation gates (e.g. $\theta$ in $R_X(\theta)$), but more generally, they can represent any tunable parameters in a quantum operation. In near-term HQC settings, the PQC is the point-of-contact between quantum and classical computational resources. That is, upon computing an objective function value based on executing these circuit runs on the quantum computer, the circuit parameters are then refined using optimization schemes on the classical computer. More recently, this model has been compared to and interpreted in the language of classical neural networks, in which the parameters of the quantum circuit are analogous to the parameters (i.e. weights, biases) of a classical neural network \cite{Killoran2018, SchuldCircuitCentric}. And just as various neural network architectures have been proposed for specific tasks, the structure of the parameterized quantum circuits can widely vary depending on the application. For instance, in the case of simulating fermionic systems, various ansatz designs, including unitary coupled-cluster \cite{McClean2016, Romero2017_2}, fermionic SWAP network \cite{Kivlichan2018}, and low-depth circuit ansatz (LDCA) \cite{DallaireDemers2018}, have been proposed. In recent years, a more heuristic but near-term approach to circuit designs, e.g. ``hardware-efficient'' circuits \cite{Kandala2017}, has been used for applications in quantum chemistry and quantum machine learning. Specifically, this circuit layout assumes a unit layer containing single-qubit operations followed by entangling two-qubit operations. In this work, we will also refer to this unit layer of gate sequence as a circuit template. This unit layer can be repeated $L$ times to provide more flexibility in the circuit. This ``multi-layer'' circuit architecture, reminiscent of deep neural networks, has been shown to provide better results in VQE \cite{Kandala2017}.

The importance of parameterized circuits has led to the development of new ansatz designs as well as studies of circuit properties and capabilities \cite{Geller2018, Du2018}. However, there remains a lack of understanding on what makes a particular parameterized circuit more powerful or useful than another. In this work, we approach this question by defining some operational quantities to characterize parameterized circuits. In the following section, we propose two descriptors, expressibility and entangling capability, that can be quantified by computing statistical properties based on sampling states from a parameterized quantum circuit template. 

\section{\label{sec:circuit_descriptors}Circuit descriptors}

In this section, we define two descriptors which we call expressibility and entangling capability, in addition to other descriptors such as the number of circuit parameters and the number of two-qubit operations. Several of these quantities or related quantities have been applied to past studies of pseudorandom quantum circuits \cite{Weinstein2008}. However, our objective in this work is not to generate pseudorandom circuits but rather study the capabilities of a PQC by quantifying its deviation from random circuits in order to approach the question of how much generalization or expressiveness is useful or enough in a PQC for a particular task.

\subsection{Expressibility}
We define expressibility as a circuit's ability to generate (pure) states that are well representative of the Hilbert space. In the case of a single qubit, this corresponds to a circuit's ability to explore the Bloch sphere. One approach for computing this notion of expressibility is to compare the distribution of states obtained from sampling the parameters of a PQC to the (expressive) uniform distribution of states, i.e. the ensemble of Haar-random states. To quantify the non-uniformity, we look to the definition of an $\epsilon$-approximate state $t$-design using the Hilbert-Schmidt norm\footnote{This definition of an approximate $t$-design employing the Hilbert-Schmidt norm can be related to alternative definitions via norm inequalities. That is, an $\epsilon$-approximate $t$-design of a particular definition and an $\epsilon'$-approximate $t$-design of another definition are related by a factor of $\text{poly}(2^{nt})$ for $n$ qubits \cite{Nakata2014}.}. That is, we are interested in computing the deviation from a state $t$-design, where $\epsilon$ may not necessarily be small.
The deviation is often quantified as:

\begin{align}
A = \int_{\textbf{Haar}} (\dyad{\psi})^{\otimes t} d\psi - \int_{\bm{\mathcal{E}}} (\dyad{\phi})^{\otimes t} \ d\phi,
\end{align}

\noindent where the first integral is taken over all pure states over the Haar measure, and the second integral is taken over all states $\ket{\phi}$ according to an ensemble in consideration, $\bm{\mathcal{E}}$. In our application, we consider a specific case of a state ensemble that is generated by uniformly sampling in the parameter space. Choosing an uninformative distribution on $\boldsymbol\theta$ is to reflect the fact that in practice, the optimizer employed in variational algorithms initially lacks knowledge of the parameter landscape.
Considering such an ensemble, $A$ can be rewritten as:

\begin{align}
A = \int_{\textbf{Haar}} (\dyad{\psi})^{\otimes t} d\psi - \int_{\boldsymbol\Theta} (\dyad{\psi_{\boldsymbol\theta}})^{\otimes t} \ d\boldsymbol\theta,
\end{align}

\noindent where the latter integral, which we call $\mu_t$, is taken over all states over the measure induced by uniformly sampling the parameters $\boldsymbol\theta$ of the PQC.
Once matrix $A$ is defined,
consider the square of its Hilbert-Schmidt norm, which can be expanded as follows:

\begin{align}
\label{eq:hs_norm_projector}
\norm{A}_{\text{HS}}^2 &= \text{tr}(A^\dag A) \\[8pt]
&= \text{tr}\Bigg( \Bigg[\frac{\Pi^t_{\text{sym}}}{d^{(t)}_{\text{sym}}} - \mu_t\Bigg]^\dag\Bigg[\frac{\Pi^t_{\text{sym}}}{d^{(t)}_{\text{sym}}} - \mu_t\Bigg] \Bigg) \\[8pt]
&= \text{tr}\Bigg[ \bigg(\frac{\Pi^t_{\text{sym}}}{d^{(t)}_{\text{sym}}}\bigg)^2 \Bigg] - 2 \ \text{tr} \Bigg( \frac{\Pi^t_{\text{sym}} \ \mu_t}{d^{(t)}_{\text{sym}}} \Bigg) + \text{tr} (\mu_t^2) \\[8pt]
&= \frac{1}{d^{(t)}_{\text{sym}}} - \frac{2}{d^{(t)}_{\text{sym}}} + \text{tr} (\mu_t^2) \\[8pt]
\label{eq:hs_projector_properties}
&= \frac{-1}{d^{(t)}_{\text{sym}}} + \text{tr} (\mu_t^2).
\end{align}

\noindent In Eq.(\ref{eq:hs_norm_projector}), we substitute the Haar integral with the normalized projector onto the symmetric subspace of $t$ $2^n$-dimensional spaces, or $\frac{\Pi^t_{\text{sym}}}{d^{(t)}_{\text{sym}}}$, where $d^{(t)}_{\text{sym}}$ denotes the dimension of the subspace. Using properties of the projector, specifically properties of a general projector as well as the expansion of the Haar integral as a linear combination of subsystem permutation operators, the expression simplifies to a sum of two terms -- a constant (provided a fixed number of qubits) plus a purity term for $\mu_t$, as shown in Eq.(\ref{eq:hs_projector_properties}). The purity term is:

\begin{align}
\text{tr}(\mu_t^2) = \int_{\boldsymbol\Theta} \int_{\boldsymbol\Phi} \abs{\braket{\psi_{\boldsymbol\theta}}{\psi_{\boldsymbol\phi}}}^{2t} \ d\boldsymbol\theta \ d\boldsymbol\phi,
\end{align}

\noindent which as seen above, is integrated twice over the parameter space. Each term in Eq.(\ref{eq:hs_projector_properties}) can be expressed as a quantity called the \emph{frame potential} \cite{Roberts2017}:

\begin{align} \label{eq:hs_norm_as_fp}
\mathcal{F}^{(t)}_{\text{Haar}} = \frac{1}{d^{(t)}_{\text{sym}}} \ \ \text{and} \ \ \mathcal{F}^{(t)} = \int_{\boldsymbol\Theta} \int_{\boldsymbol\Phi} \abs{\braket{\psi_{\boldsymbol\theta}}{\psi_{\boldsymbol\phi}}}^{2t} \ d\boldsymbol\theta \ d\boldsymbol\phi,
\end{align}

\noindent
where the generalized $t$-th frame potential $\mathcal{F}^{(t)}$ for the state $\ket{0} = \ket{0}^{\otimes n}$ is defined as: 

\begin{align}
\mathcal{F}^{(t)}_{\ketbra{0}} &= \int d\mu(U) \ d\mu(V) \ [\bra{0} U V^\dag \ket{0}\bra{0} V U^\dag \ket{0}]^t \\[6pt]
&= \int d\mu(\ket{\psi}) \ d\mu(\ket{\phi}) \ \abs{\bra{\psi}\ket{\phi}}^{2t}.
\end{align}

\noindent In this definition, the measures $\mu$ correspond, respectively, to the distribution over unitaries and the distribution over states induced by applying unitaries from the distribution to the initial state $\ket{0}$.
For an ensemble of Haar random states or a state $t$-design, $\mathcal{F}^{(t)} = \frac{1}{d^t_{\text{sym}}} = \frac{(t)!(N-1)!}{(t+N-1)!}$, where $N=2^n$ for $n$ qubits. These frame potential values 
saturate the, so-called,
\emph{Welch bounds} \cite{Datta2012}. Past studies have shown that 
the first
$t$ frame potentials of an ensemble 
achieve their minimum values 
(Welch bounds) if and only if the ensemble is a state $t$-design \cite{Renes2004, Klappenecker2005, Bengtsson2017}, motivating frame potentials as  ``probe[s] of randomness'' \cite{Roberts2017}. This lower bound can be observed by noting that $\norm{A}_{\text{HS}}^2 \geq 0$, and thus Eq.(\ref{eq:hs_norm_as_fp}) implies $\mathcal{F}^{(t)} \geq \mathcal{F}^{(t)}_{\text{Haar}}$. Given this property, one potential approach to quantifying expressibility could be to estimate the first few frame potentials using sampled parameterized states and comparing them against the Haar values. However, using this method, it is difficult to derive a single meaningful score. Instead, we 
draw inspiration from the structure of
frame potentials to 
establish an operationally meaningful score of expressibility.

\subsubsection{\label{subsec:estimate_expr}Estimating expressibility}
Frame potentials can be understood as the $t$-th moments of the distribution of state overlaps:

\begin{equation}
\textup{p}(F=\abs{\braket{\psi_{\boldsymbol\theta}}{\psi_{\boldsymbol\phi}}}^2).
\end{equation}

\noindent This is seen as follows:
\begin{align}
\mathcal{F}^{(t)} &= \int_{\boldsymbol\Theta} \int_{\boldsymbol\Phi} \abs{\braket{\psi_{\boldsymbol\theta}}{\psi_{\boldsymbol\phi}}}^{2t} \ d\boldsymbol\theta \ d\boldsymbol\phi \\[8pt]
&= \mathbb{E}_{\boldsymbol\theta} \ \mathbb{E}_{\boldsymbol\phi} [(\abs{\braket{\psi_{\boldsymbol\theta}}{\psi_{\boldsymbol\phi}}}^{2})^t]\\[8pt]
&= \mathop{\mathbb{E}} [F^t], \ \ F = \abs{\braket{\psi_{\boldsymbol\theta}}{\psi_{\boldsymbol\phi}}}^{2}.
\end{align}

\noindent For the ensemble of Haar random states, the analytical form of the probability density function of fidelities is known: $P_{\text{Haar}}(F) = (N-1)(1-F)^{N-2}$, where $F$ corresponds to the fidelity and $N$ is the dimension of the Hilbert space \cite{Zyczkowski2005}. To probe the non-uniformity of the set of states generated by a PQC, we propose comparing the resulting distribution of state fidelities generated by the sampled ensemble of parameterized states to that of the ensemble of Haar random states.
In practice, we can estimate the fidelity distributions or estimate the first few moments of fidelity $F$ by independently sampling pairs of states (i.e. sampling pairs of parameter vectors and obtaining parameterized states) and treating the corresponding fidelities as random variables.
Using this sampling technique, we can show that the resulting sample mean is an unbiased estimator of the population mean.
After collecting sufficient samples of state fidelities\footnote{Refer to Appendix \ref{app:n_samples} for a discussion on the sample size.}, the Kullback-Leibler (KL) divergence \cite{Kullback1951}, often used in machine learning applications, between the estimated fidelity distribution and that of the Haar-distributed ensemble can be computed to quantify expressibility (``Expr''):

\begin{align}
\text{Expr} = \kldiv{\hat{P}_{\text{PQC}}(F; \boldsymbol\theta)}{P_{\text{Haar}}(F)},
\end{align}

\noindent where $\hat{P}_{\text{PQC}}(F; \boldsymbol\theta)$ is the estimated probability distribution of fidelities resulting from sampling states from a PQC. Due to a finite sample size, the probability distribution is estimated with a histogram. Therefore, we must choose a discretization of the two probability distributions in order to numerically estimate the KL divergence. 
By this scoring method, a PQC with a resulting fidelity distribution that corresponds to a lower KL divergence with respect to that of Haar, is a more expressible circuit. In the least-expressible case, in which a circuit has fixed gates, e.g. $I$ (versus parameterized gates), that always outputs the same state, the upper bound expressibility value is $(N-1) \ \text{ln}(n_{\text{bin}})$ for system size of $N$ and number of bins for the histograms, $n_{\text{bin}}$.

Defining expressibility in terms of the KL divergence grants us an operational meaning for this numerical value: expressibility is the amount of information that is lost if we were to approximate the distribution of state fidelities generated by a PQC using that of Haar random states.
We expect that, for well-behaved distributions, the expressibility will be zero if and only if the distribution of states is the Haar distribution. However, a proof of this fact has so far eluded us.

\subsubsection{Expressibility: single qubit demonstration}

% One-qubit example
\begin{figure*}
\centering
\includegraphics[width=0.85\textwidth]{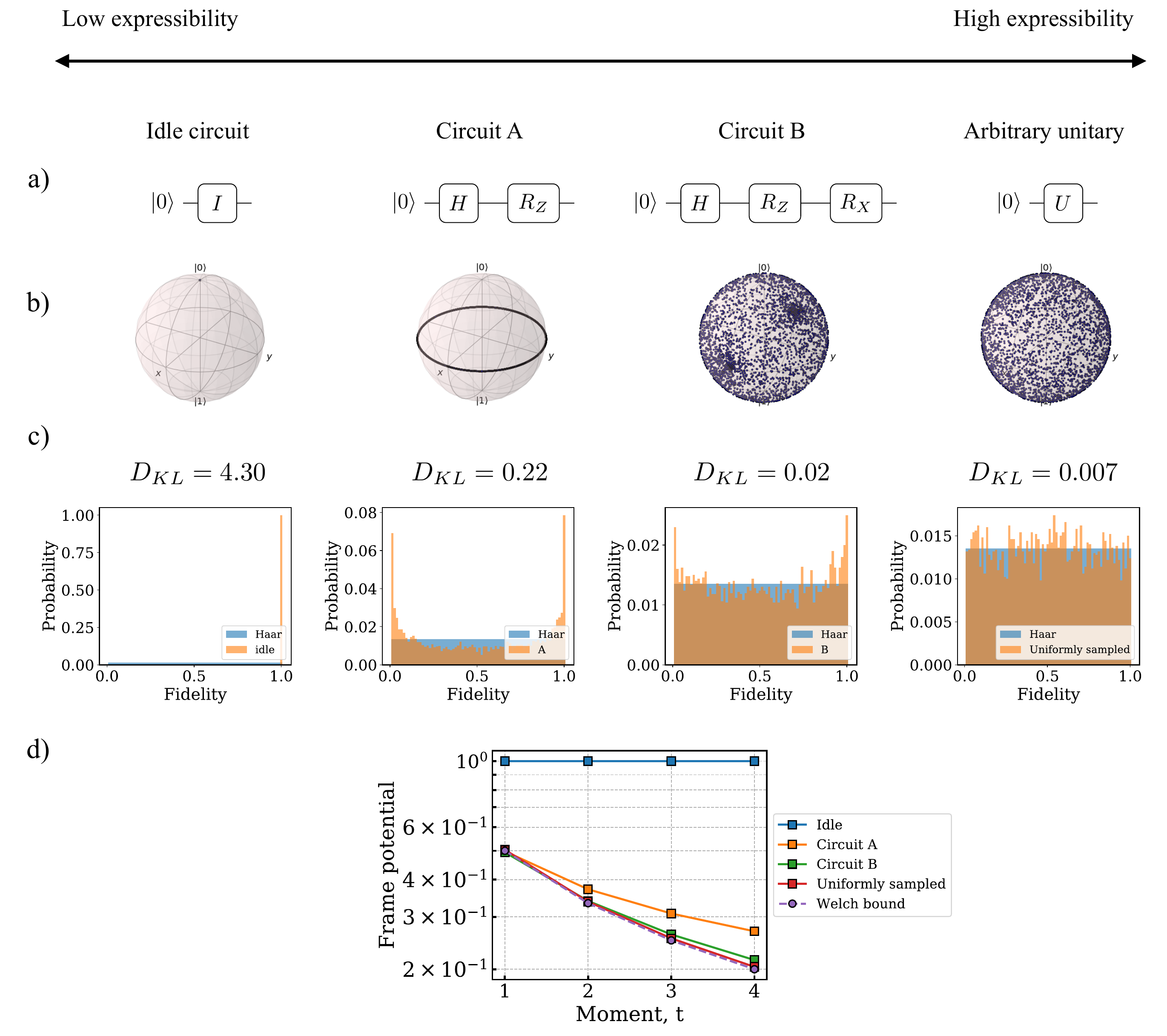}
\caption{Quantifying expressibility for single-qubit circuits. (a) Circuit diagrams are shown for the four types of circuits. (b) For each circuit, $1000$ sample pairs of circuit parameter vectors were uniformly drawn, corresponding to $2000$ parameterized states that were plotted on the Bloch sphere using \texttt{QuTiP} \cite{qutip}. (c) Histograms of estimated fidelities are shown, overlaid with fidelities of the Haar-distributed ensemble, with the computed KL divergences reported above the histograms. (d) The frame potential estimates for the first four moments are plotted for each circuit, with the Haar values (Welch bounds) plotted using a purple dotted line.} 
\label{fig:single_qubit_example}
\end{figure*}

We describe a simple example of computing expressibility to clarify its construction.
Consider the single-qubit circuits shown in Fig. \ref{fig:single_qubit_example}a that ranges in their abilities to explore the Bloch sphere. The ``idle'' circuit is an un-expressive circuit that consists of an identity or idle gate acting on the reference state $\ket{0}$. The states obtained from uniformly sampling $\boldsymbol\theta$ (rotation about the $Z$-axis) on circuit $A$ are limited to an exploration about the equator of the Bloch sphere. The set of states generated by circuit $B$, however, is expected to have a better coverage of the Bloch sphere due to the additional degree of freedom provided by the $X$-rotation. Finally, as a reference, we uniformly sample single-qubit unitary matrices to simulate the most expressible circuit case. Fig. \ref{fig:single_qubit_example}b displays simulated data of sampled states ($2000$ points) plotted on the Bloch sphere. Because we are sampling uniformly in the parameter space (versus the state space), sampling states from circuit $B$ will lead to greater concentrations of states in the $+x$ and $-x$ poles. In Fig. \ref{fig:single_qubit_example}c, the estimated histograms of fidelities are displayed for each circuit, overlaid with the histogram of fidelities for the Haar-distributed ensemble for comparison. A bin 
number
of $75$ was used to generate the histograms.
Although the KL divergences will vary with different bin number, we expect the observations coming from relative quantitative comparisons among circuits to remain the same.
Above each histogram, the KL divergences are reported to quantify the deviation, in which a lower KL divergence value corresponds to more favorable expressibility or a greater closeness to Haar random states. For the least expressible case, i.e. the idle circuit, the expected KL divergence is $\text{ln}(75) \approx 4.3$. In Fig. \ref{fig:single_qubit_example}d, the estimated frame potentials for the first four moments are plotted, in which we see that circuits that are expected to be more expressible have frame potential values that are lower or closer to the corresponding values of the Haar-distributed ensemble.

\subsection{\label{sec:entangling_capability}Entangling capability}
In the context of variational HQC algorithms, potential advantages of generating highly entangled states with low-depth circuits include the ability to efficiently represent the solution space for tasks such as ground state preparation or data classification, and to capture non-trivial correlation in the quantum data \cite{SchuldCircuitCentric, Kandala2017}. Accordingly, in VQE \cite{Kandala2017} and quantum machine learning \cite{SchuldCircuitCentric, Havlicek2018}, strongly entangling circuits have been realized by appending and repeating layers comprised of varying configurations of two-qubit gates, e.g. CNOT, CZ, and their parameterized variants. Though the powers and advantages of such ``entanglers'' have been empirically demonstrated for specific problems, we propose computing the Meyer-Wallach (MW) entanglement measure \cite{Meyer2002} as a way to quantify the entangling capability of a parameterized quantum circuit, or its ability to generate entangled states, independent of the problem at hand. The Meyer-Wallach entanglement measure, often denoted as $Q$, is a global measure of multi-particle entanglement for pure states. While there exist several methods for quantifying entanglement as a resource \cite{Nielsen2003}, the MW measure was chosen due to its scalability and ease of computation.

\subsubsection{Meyer-Wallach measure}

The Meyer-Wallach entanglement measure, or $Q$, is defined as follows. For a system of $n$ qubits, consider a linear mapping $\iota_j(b)$ that acts on the computational basis:

\begin{equation} \label{eq:partial_trace_mapping}
\begin{split}
\iota_j(b) \ket{b_1 ... b_n} =  \delta_{bb_j} \kket{b_1 ... \hat{b}_j ... b_n}
\end{split}
\end{equation}

\noindent where $b_j \in \{0, 1\}$ and the symbol $ \ \hat{} \ $ denotes absence of the $j$-th qubit. Meyer and Wallach define the entanglement measure $Q$ as:

\begin{equation} \label{eq:meyer_wallach}
\begin{split}
Q(\ket{\psi}) \equiv \frac{4}{n} \sum_{j=1}^n D \big( \iota_j(0) \ket{\psi}, \iota_j(1) \ket{\psi} \big),
\end{split}
\end{equation}

\noindent where the generalized distance $D$ is:

\begin{equation} \label{eq:meyer_wallach_distance}
\begin{split}
D(\ket{u}, \ket{v}) = \frac{1}{2} \sum_{i,j} \abs{u_i v_j - u_j v_i }^2,
\end{split}
\end{equation}

\noindent with $\ket{u} = \sum u_i \ket{i}$ and $\ket{v} = \sum v_i \ket{i}$. $D$ can be understood as the square of the area of the parallelogram created by vectors $\ket{u}$ and $\ket{v}$.
By construction, $Q$ has the following properties: (1) $Q$ is invariant under local unitaries, (2) $0 \leq Q \leq 1$, and (3) $Q(\ket{\psi}) = 0$ if and only if $\ket{\psi}$ is a product state. 
For instance, $Q(\ket{01}) = 0$, while $Q(\frac{\ket{00} + \ket{11}}{\sqrt{2}}) = 1$.
It can be shown that $Q$ is the average linear entropy (i.e. $1-\tr{\rho^2}$) of all the single qubit reduced states  \cite{Brennen2003}.
A drawback of the MW measure is that it is fairly undiscerning with respect to different types of entanglement. As described in \cite{Brennen2003}, for instance, the state $\ket{\Psi} = \frac{\ket{00}_{1,2} + \ket{11}_{1,2}}{\sqrt{2}} \otimes \frac{\ket{00}_{3,4} + \ket{11}_{3,4}}{\sqrt{2}}$ and the four-qubit GHZ state both correspond to the maximum $Q$ value of 1. However, alternative measures of entanglement, such as the Schmidt number \cite{Terhal2000}, rank the GHZ state as having higher entanglement than $\ket{\Psi}$.
Nevertheless, we chose the MW measure because it has been used as an effective probe of entanglement for a wide range of applications in quantum information, including characterization of entangled states involved in quantum error correcting codes \cite{Meyer2002} and quantum phase transitions \cite{Somma2004}. The MW measure has also been applied as a tool for dynamically tracking the convergence of pseudorandom circuits by computing the deviation of the MW measure from the Haar value \cite{Weinstein2008}, and the evolution of entanglement for an instance of Grover's algorithm \cite{Meyer2002}. Over the course of a variational HQC algorithm, parameters of a PQC are dynamically tuned to reach the solution space for a given problem. Therefore, the MW measure seems to be especially well-suited for quantifying the entangling capability of a parameterized quantum circuit by quantifying the number and types of entangled states it can generate.

\subsubsection{Estimating entangling capability}

We define the entangling capability of a parameterized quantum circuit as the average Meyer-Wallach entanglement of states generated by the circuit. For a given PQC, we can estimate this value by sampling the circuit parameters and computing the sample average of the MW measure of output states. More precisely, we take the estimate of the entangling capability to be:

\begin{equation} \label{eq:entangling_capability_score}
\begin{split}
\text{Ent} = \frac{1}{\abs{S}} \sum_{\boldsymbol\theta_i \in S} Q \big( \ket{\psi_{\boldsymbol\theta_i}} \big),
\end{split}
\end{equation}

\noindent where $S= \{ \boldsymbol\theta_i \}$ is the set of sampled circuit parameter vectors. Using this measure, a parameterized circuit that outputs only product states will have an entangling capability score of $0$, whereas one that always produces highly entangled states will correspond to a score close to $1$. For cases in between, the mean will lie in between the two values, and for a closer investigation to understand to what extent the circuit is able to produce entangled states, one can compute statistical properties of the sample distribution of $Q$ values. Note that we can recycle the states generated in the expressibility study for estimating this entangling capability.

\subsection{Circuit costs}

In order to establish a fair comparison of the expressibility and entangling capability among circuits, the ``costs'' of implementing the various circuits should also be taken into account. The three costs we consider are the circuit depth, circuit connectivity, and number of parameters. The first two costs are particularly relevant for NISQ computers, which are limited in their coherence times and connectivity among the qubits. In addition to the circuit depth, we also track the number of two-qubit gates for each circuit to estimate the difficulty of implementing the circuit on a quantum device. That is, we expect a circuit comprising significantly of costly two-qubit gates to execute with a lower program fidelity. For the circuits we consider in the study, the configuration of their two-qubit gates determine the overall connectivity, where the use of non-local two-qubit gates generally increases the complexity of the required qubit topology. In principle, one can decompose the circuit to map onto alternative topologies, but the resulting circuit may incur a high cost in the circuit depth and number of gates. Therefore, we consider three possible configurations of two-qubit gates in our circuits: nearest-neighbor, ring topology, and all-to-all connectivities. The number of parameters is especially relevant for variational quantum algorithms, in which an optimization routine on the classical computer is tasked with updating and refining the parameter values. Thus, we use the number of parameters as a rough measure for the difficulty of optimization.

\section{\label{sec:numerical}Numerical experiments}
In this section, we demonstrate the use of the descriptors for characterizing parameterized quantum circuits by computing them for a select set of circuits composed of different configurations of single-qubit and two-qubit gate operations. Parameterized gates used in this work are $R_X$, $R_Y$, and $R_Z$, defined in \cite{NielsenChuang}. We set the circuit width to four qubits, but for an analysis of these circuit templates at larger widths, the reader should refer to Appendix \ref{app:circuit_width_analysis}. Circuit simulations presented in this work are implemented using the Forest platform \cite{quil}.

\begin{figure*}
\centering
\includegraphics[height=0.85\textheight]{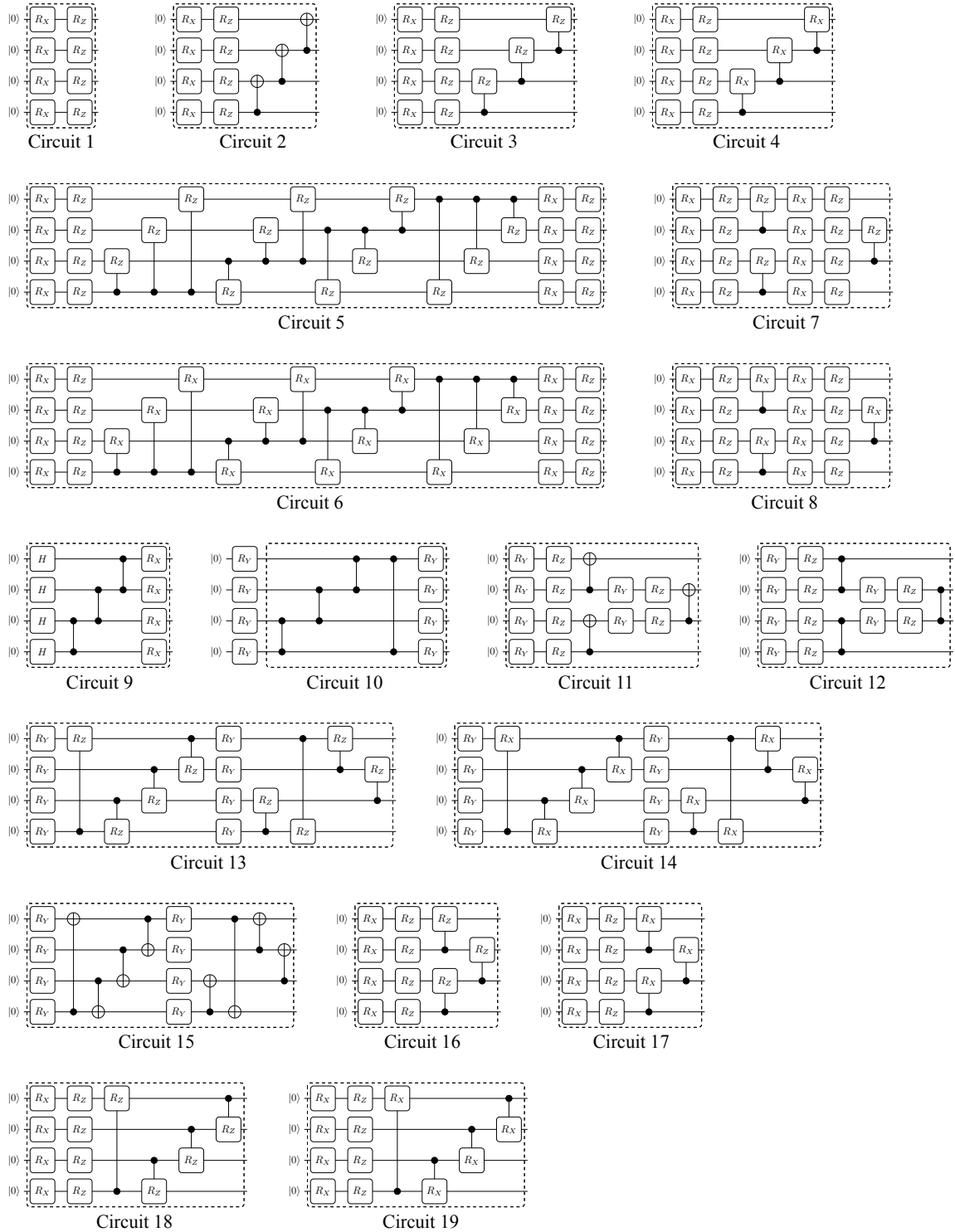}
\caption{A set of circuit templates considered in the study, each labeled with a circuit ID. The dashed box indicates a single circuit layer, denoted by $L$ in the text, that can be repeated. Gates $R_X$, $R_Y$, and $R_Z$ are parameterized. Several circuit templates are from or inspired by past studies. Circuit diagrams were generated using $\texttt{qpic}$ \cite{qpic}.}
\label{fig:test_circuits}
\end{figure*}

Several circuit designs in Fig. \ref{fig:test_circuits} were derived or inspired by past studies. For instance, circuits $5$ and $6$ were developed in \cite{Sousa2006} as programmable quantum circuits and were applied to train the quantum autoencoder in  \cite{Romero2017}. Circuits $7$ and $8$ were used as encoding and un-encoding circuits for the QVECTOR algorithm \cite{Johnson2017}. Circuit $9$ was a Quantum Kitchen Sinks ansatz considered in \cite{Wilson2018}. Circuit $10$ followed the hardware-efficient circuit architecture from \cite{Kandala2017}. Circuit $11$ and $12$ were Josephson sampler circuits from \cite{Geller2018}. Circuits 13-15 and 18-19 followed the circuit-block construction from \cite{SchuldCircuitCentric} used for data classification.

In the following subsections, we review observations from our simulations for expressibility and entangling capability. For an in-depth discussion of trends e.g. in gate types and configurations, the reader should refer to Section \ref{sec:discussion}.

% Figure of test circuits

\subsection{\label{sec:test_circuit_expressibility}Expressibility observations}

% Figure of 4q expressibility scores
\begin{figure*}
\centering
\includegraphics[width=0.9\textwidth]{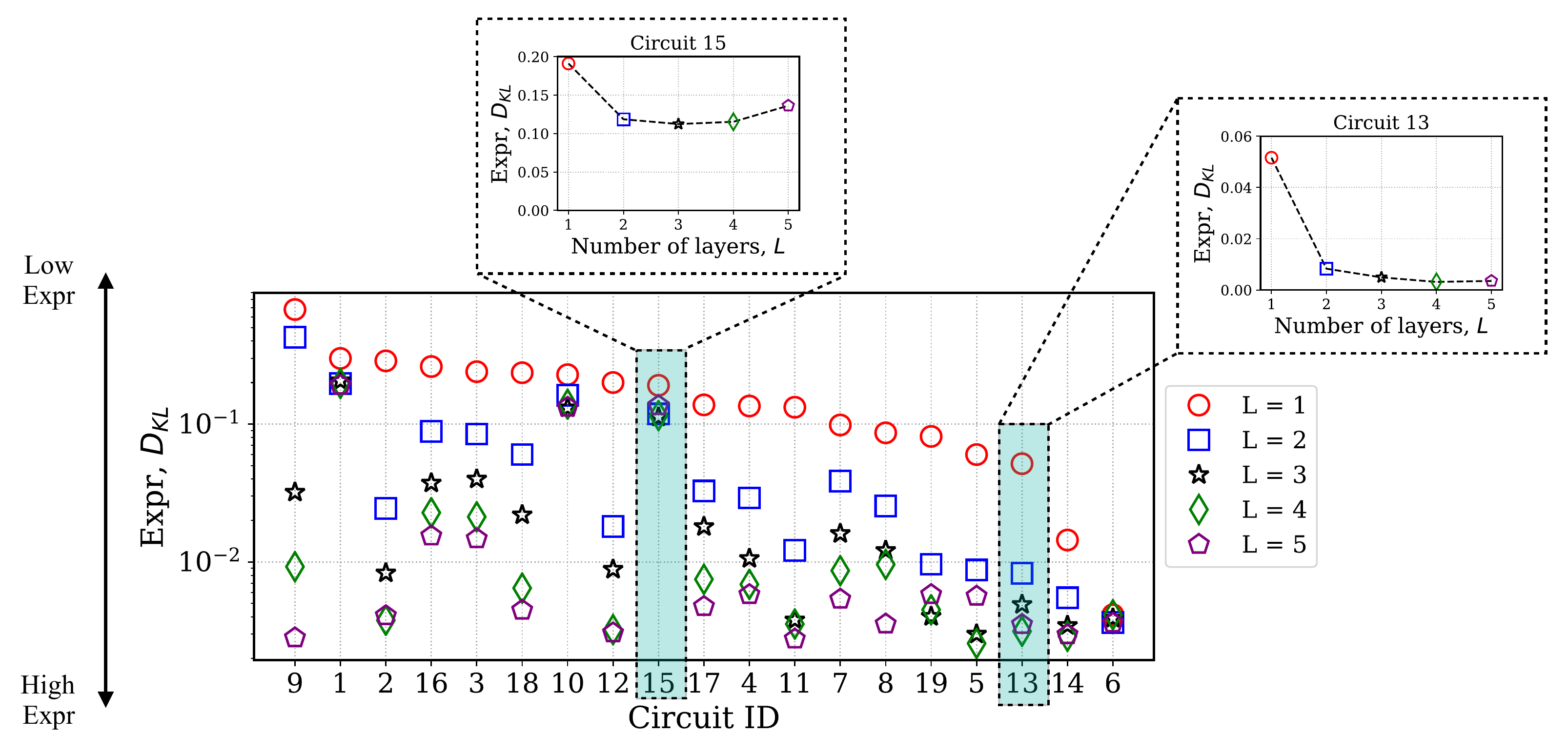}
\caption{
Expressibility values (or KL divergences) computed for the benchmark circuits from Fig. \ref{fig:test_circuits} with circuit widths of $n=4$ qubits. Marker colors indicate different numbers of circuit layers ($L$) applied to a circuit template. Data for each circuit are presented in the order of increasing expressibility (i.e. decreasing KL divergence) for $L=1$. The zoomed-in plots of the two highlighted regions show instances of ``expressibility saturation'' discussed in the text.}
\label{fig:exp_score_4q}
\end{figure*}

One of the main advantages of employing a multi-layered parameterized circuit to variational HQC algorithms is the potential to systematically extend its ``flexibility,'' or the ability to represent a wider class of states, simply by adding more layers to the original circuit, which was demonstrated in the Appendix of Ref. \cite{Kandala2017}. In this section, we provide a deeper analysis into the effects of multi-layering by computing how much each layer contributes to the overall expressibility for a given circuit template. We numerically demonstrate that,
while expressibility generally improves with increased circuit layers, it does so differently depending on the circuit template.
This implies that, provided two circuits with different expressibility values, it may be preferred (e.g. more economic in depth and gate count) to select the less expressible circuit template but with added layers, in the case that multi-layering boosts the expressibility of that circuit.

To compute and compare the expressibility among parameterized circuit templates in Fig. \ref{fig:test_circuits}, 5000 state fidelities were sampled for each circuit instance to construct its histogram, using a bin size of 75. For each circuit template, we considered instances in which the unit layer was repeated up to five times, i.e. $L_{\text{max}}=5$. Expressibility for these circuit instances are computed and plotted in Fig. \ref{fig:exp_score_4q}, in which the data is organized such that circuits are ordered based on ascending values of expressibility for the $L=1$ instance. This ordering scheme enables us to track the changes in the expressibility values as well as changes in the relative ranking (by expressibility) with increased values of $L$. 

We expect that circuits with more gates will generally have a more favorable expressibility. In Appendix \ref{app:exp_saturation} we compare expressibility among circuits as a function of the number of two-qubit gates.
Expressibility as a function of two-qubit gate number may be a more fair way to compare two circuit templates.
Two-qubit gates are more costly in terms of time required for implementation as well as the noise they contribute.
However, care must be taken with such a cost analysis.
In practice, the circuits proposed must be transpiled into the gate set that is native to the particular device.
Depending on the device, the resulting total number of two-qubit gates will be different.
For now, we will consider expressibility
as a function of circuit layer number, and reserve the expressibility per-two-qubit-gate comparisons for Appendix \ref{app:exp_saturation}.

We first grouped circuits based on closeness in expressibility values at $L=1$. This enables us to make comparisons among circuits that start with similar expressibility values. In the following, we describe observations on the groups of circuits, before extending the analysis to $L>1$. At $L=1$:

\begin{itemize}
    \item Circuit 9 is the least expressible of the nineteen circuit templates, i.e. corresponds to the highest KL divergence of approximately $0.68$. By construction, a single layer of circuit 9 can produce high-entanglement ($Q$=1) states but cannot efficiently explore low-entanglement states. This results in an unfavorable expressibility value.
    \item Circuits 1, 2, 16, 3, 18, 10, 12, and 15 had
    comparable
    expressibility values of around 0.2.\footnote{Circuits 3 and 16 are in fact equivalent, and we chose to include both circuits to reflect the fact that parameterized circuits are often generated by choosing a particular stencil (fixed placement of arbitrary gates), and then populating the gate slots from a selected gate set. In principle, we should be able to (and do) capture the circuit equivalence from matching descriptor values. Moreover, these circuits may not yield identical behavior in an experimental setting.} Circuits 1-4 were originally devised to demonstrate a systematic increase in expressibility, as (fixed) two-qubit gates are added to circuit 1 to construct circuit 2. From circuit 2, the CNOTs are replaced with parametric two-qubit gates to construct circuits 3 and 4. Using a single circuit layer, there was no significant increase in expressibility for circuit 2 and 3, compared to that of circuit 1. However, a higher expressibility was observed for circuit 4.
    \item Circuits 17, 4, and 11 achieved expressibility values near 0.09. These circuits, which employed two-qubit gates in a nearest-neighbor fashion, were more expressible than circuit 15 that employed two-qubit gates in a ring topology. This may be due to the parametric two-qubit gates used by the three circuits, compared to the (static) CNOTs used in circuit 15.
    \item Circuits 7, 8, and 19 achieved expressibility values near 0.09. Circuit 8 can be viewed as a parameterized variant of circuit 11, where the additional degrees of freedom from the extended parameterization led to a (small) improvement in expressibility. 
    \item Circuit templates exhibiting favorable expressibility ($D_{\text{KL}} < 0.02$) included circuits 5, 13, 14, and 6, in ascending order. Despite the use of costly all-to-all configurations of two-qubit gates employed by circuit 5, circuit 13, which employs the circuit-block construction (ranges of control set to 1 and 3), achieved more favorable expressibility. 
    \item Circuit 6 was the most expressible circuit at $L=1$, achieving a low KL divergence with a single circuit layer. However, even a single layer of circuit 6 requires a high circuit depth and number of parameters.
\end{itemize}

For each circuit template, its unit layer was repeated $L$ times, leading to a general increase in expressibility. However, shifts in the ordering of circuits by expressibility were observed, as shown in Fig. \ref{fig:exp_score_4q} for $L>1$ instances. For example, two layers of circuit 2 are more expressible than two layers of circuits 16 or 3. 
This implies that the rates of change in expressibility vary among the circuits. 
For fourteen out of the nineteen circuit templates that were considered,
expressibility of
these circuits at $L=2$
increased (i.e. their corresponding KL divergences decrease)
by more than $60\%$ of their respective values at $L=1$. From $L=2$ to $L=3$,
expressibility %\hs{[KL divergence]}
values of six circuits, including that of circuit 9,
increased
by more than $60\%$. That is, differences in circuit structures corresponded to having different rates of increase in
expressibility from layer to layer.
One consequence is that there may be a circuit template that corresponds to a less favorable expressibility with limited layers but reaches a significantly better value with sufficient layers. 
For example, with just three layers the expressibility of circuit 11 ``catches up'' to that of circuit 6 and does so with a simpler circuit connectivity.
Finally, we observe that the 
value of expressibility 
for increasing
layers 
appears to ``saturates'' at different values for different circuit templates.
We further explore and elaborate on 
these observed behaviors
in Section \ref{sec:discussion}.

\subsection{\label{sec:test_circuit_entangling_capability}Entangling capability observations}

% Entangling capability scores
\begin{figure}
\centering
\includegraphics[width=0.49\textwidth]{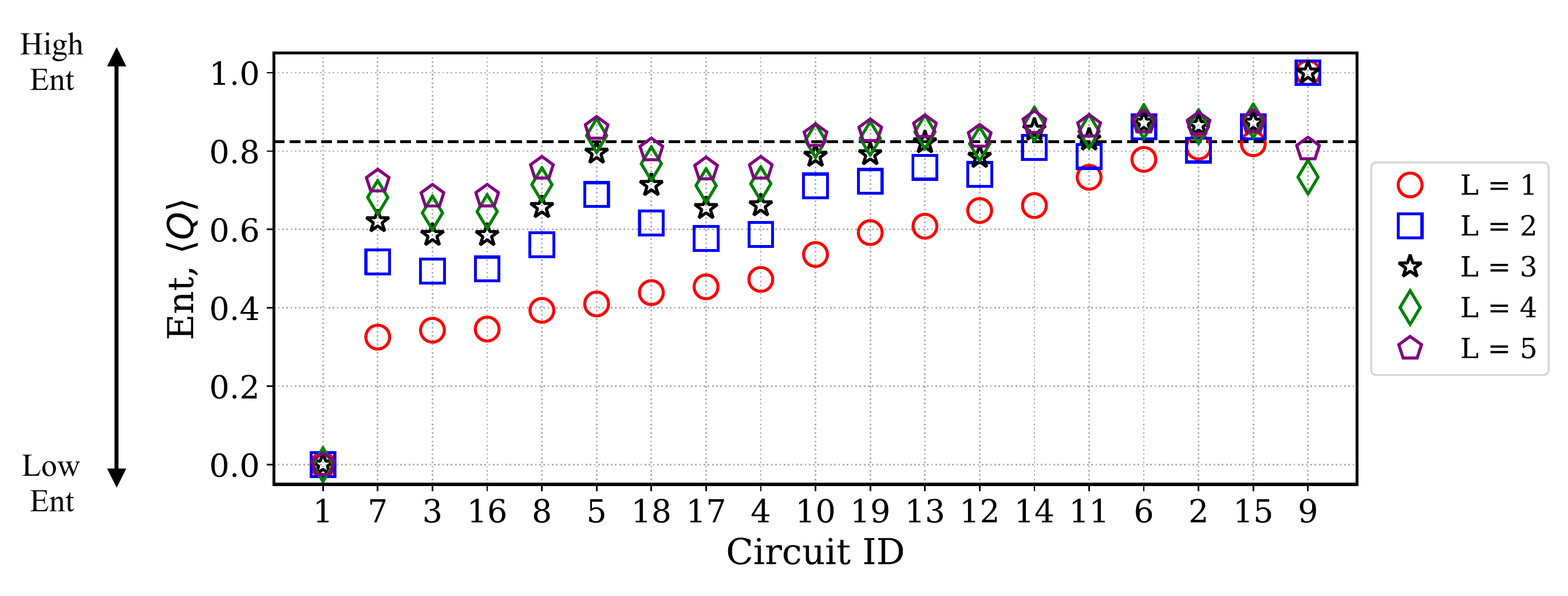}
\caption{Entangling capability values for the benchmark circuits with widths of $n=4$. Marker colors indicate the different numbers of circuit layers ($L$) applied. Data for each circuit are presented in the order of increasing entangling capability for $L=1$. The black dashed line shows the mean $Q$ value for random pure states.}
\label{fig:entangling_capability_test_circuits}
\end{figure}

As noted earlier, the entangling capability, or the average $Q$, can be estimated by recycling the set of parameterized states used to estimate the expressibility. We perform this computation for the circuit templates in Fig. \ref{fig:test_circuits}, this time exploring the effects of multi-layering in generating high-entanglement states.
The two ``extreme'' cases are circuit 1 and 9, in which the former is composed only of local unitaries and thus all the states sampled by tuning its parameters have $Q$ values of 0, as observed in the figure. By contrast, circuit 9 outputs high-entanglement states, resulting in $Q$ values near 1. However, when up to four and five layers were added to circuit 9, a decrease in the mean was observed. At that point, the circuit is flexible enough to produce lower-entanglement states, leading to lower average $Q$. More specifically, the entangling capability for $L=4$ is lower than that for $L=5$. This 
may be explained as
an oscillatory convergence to the mean $Q$ value for Haar-random pure states:

\begin{equation} \label{eq:q_mean_random_pure_states}
\begin{split}
\langle Q \rangle_{\text{Haar}} = \frac{N − 2}{N + 1},
\end{split}
\end{equation}

\noindent given an $N$-dimensional Hilbert space \cite{Weinstein2008}. This alternative metric for quantifying the closeness to Haar is verified in the expressibility data for circuit $9$ in Fig. \ref{fig:exp_score_4q}, in which its expressibility monotonically increases with the number of circuit layers. For other circuits, their mean $Q$ values are distributed between 0 and 1, where their values appear to approach the theoretical average for random pure states as circuit layers are added. 
These simulations allow us to compare expressibility and entangling capability.
While a favorable expressibility corresponds to an entangling capability that approaches or converges to $\langle Q \rangle_{\text{Haar}}$, the converse is not generally true. For example, a single layer of circuit 2 
has a relatively unfavorable expressibility value, despite having a mean $Q$ value close to $\langle Q \rangle_{\text{Haar}}$.
This joint information in fact provides some insight into the \emph{types} of states explored by circuit 2, in which sampling states from this circuit allows for a ``selective'' exploration of some highly-entangled states in the Hilbert space. 

\subsection{\label{sec:test_circuit_cost}Cost estimate observations}

For each circuit template, we considered the circuit depth, number and topology of two-qubit gates, and the number of parameters to estimate the cost associated with implementing the circuit. The number of parameters and two-qubit gates and the circuit depth are shown in Table \ref{table:descriptor_values} in terms of the number of qubits $n$ and number of circuit layers $L$.
For circuit templates that employ the circuit-block construction from ref. \cite{SchuldCircuitCentric}, they are also defined by their \emph{ranges of control}, or the number of qubits between the control and target qubits to quantify the non-locality of two-qubit gates in the block.
Circuits 13-15 are each comprised of two consecutive circuit-blocks, with ranges of control set to 1 and 3, respectively. Circuits 18-19 each utilize a single circuit-block with range of control set to 1.
To compare circuits based on the cost, we group them based on the scaling in terms of the qubit number. The circuit depths for circuits 5 and 6 grow quadratically with the qubit number. Depths for circuits 2-4, 9, 10, 13-15, 18, 19 have a linear dependence with respect to the number of qubits, while the remaining circuits have a constant dependence. Similarly, for the number of two-qubit gates and circuit parameters, circuits 5 and 6 again have a quadratic dependence on the qubit number, while the remaining circuits have a linear dependence. Circuits 5 and 6 are consistently the two most costly circuits, in terms of the circuit depth and numbers of two-qubit gates and parameters.

Circuit templates in Fig. \ref{fig:test_circuits} vary in qubit topologies. For example, the nearest-neighbor two-qubit interactions in circuits 2-4, 7-9, 11, 12, 16, and 17 enable them to be mapped naturally onto a linear array of qubits. Several circuit templates employ non-local two-qubit operations to represent circuits tailored to quantum hardware that can support 
such interactions between 
qubits. Circuits 10, 18, and 19 can be executed on a ring topology (i.e. the range on control is 1 for these circuits). Circuits 13, 14, and 15 each are comprised of two circuit blocks, with ranges of control set to 1 and 3 respectively. Consequently, the overall qubit connectivity is somewhere between a ring topology and a fully connected graph topology. Lastly, circuits 5 and 6 are composed of two-qubit interactions that require all-to-all connectivity.

From the analysis of circuit costs, it was evident that, while circuits 5 and 6 were shown to exhibit favorable expressibility and entangling capability ( especially with additional circuit layers), 
this comes
at the cost of having a large number of parameters and two-qubit gates, as well as high circuit depth and qubit connectivity. These factors make the circuits less favorable for use on NISQ devices. Some promising alternatives include circuits 11, 12, and 19 that become comparable in values of expressibility and entangling capability with additional layers, as those of circuits 5 and 6 while also maintaining reasonable circuit costs.

% Circuit costs for 19 circuits
\begin{table*}[ht]
\def\arraystretch{1.3}\tabcolsep=5pt
\begin{tabular}{|c|c|c|c|}
\hline
\textbf{Circuit ID} & \textbf{\begin{tabular}[c]{@{}c@{}}Number of\\ parameters\end{tabular}} & \textbf{\begin{tabular}[c]{@{}c@{}}Number of\\ two-qubit gates\end{tabular}} & \textbf{Circuit depth} \\ \hline
1   & $2nL$           &  $0$         & $2L$         \\ \hline
2   & $2nL$           &  $(n-1)L$    & $(n+1)L$     \\ \hline
3   & $(3n-1)L$       &  $(n-1)L$    & $(n+1)L$     \\ \hline
4   & $(3n-1)L$       &  $(n-1)L$    & $(n+1)L$     \\ \hline
5   &  $(n^2 + 3n)L$  &  $(n^2-n)L$  & $(n^2-n+4)L$ \\ \hline
6   &  $(n^2 + 3n)L$  &  $(n^2-n)L$  & $(n^2-n+4)L$ \\ \hline
7   &  $(5n-1)L$      &  $(n-1)L$    & $6L$         \\ \hline
8   &  $(5n-1)L$      &  $(n-1)L$    & $6L$         \\ \hline
9   &  $nL$           &  $(n-1)L$    & $(n+1)L$     \\ \hline
10  &  $n+nL$         &  $nL$        & $1+(n+1)L$   \\ \hline
11  &  $(4n-4)L$      &  $(n-1)L$    & $6L$         \\ \hline
12  &  $(4n-4)L$      &  $(n-1)L$    & $6L$         \\ \hline
13  &  $\big( 3n + \frac{n}{\text{gcd}(n, 3)}  \big) L$  &  $\big( n + \frac{n}{\text{gcd}(n, 3)} \big)L$ &  $ \big( 2 + n + \frac{n}{\text{gcd}(n, 3)}  \big) L$   \\ \hline
14  &  $\big( 3n + \frac{n}{\text{gcd}(n, 3)}  \big) L$  &  $\big( n + \frac{n}{\text{gcd}(n, 3)} \big)L$ &  $ \big( 2 + n + \frac{n}{\text{gcd}(n, 3)}  \big) L$   \\ \hline
15  &  $2nL$  &  $\big( n + \frac{n}{\text{gcd}(n, 3)} \big)L$ &  $ \big( 2 + n + \frac{n}{\text{gcd}(n, 3)}  \big) L$    \\ \hline
16  & $(3n-1)L$       &  $(n-1)L$  & $4L$   \\ \hline
17  & $(3n-1)L$       &  $(n-1)L$  & $4L$  \\ \hline
18  & $3nL$  &  $nL$ & $(n+2) L$ \\ \hline
19  & $3nL$  &  $nL$ & $(n+2) L$ \\ \hline
\end{tabular}
\caption{Cost estimates for circuits from Fig. \ref{fig:test_circuits}, i.e. the number of parameters, number of two-qubit operations, and circuit depth in terms of $n$, number of qubits and $L$, number of circuit layers.}
\label{table:descriptor_values}
\end{table*}

\section{\label{sec:discussion}Discussion}
With the statistical properties and cost estimates computed for the set of test circuits, we discuss several observations and trends noted in the simulation data that may serve as guides for designing new or improved parameterized quantum circuits for variational HQC algorithms.

\bigbreak

\noindent \textbf{{Expressibility saturation}}.
As additional layers are added to parameterized quantum circuits, the expressibility value does not always continue to improve. For each of the circuits studied, there is a layer number beyond which the expressibility of added layers ``saturates.'' An example of the phenomenon is shown in the two call-outs of Fig. \ref{fig:exp_score_4q}. 
Circuit 15 saturates to a value around 0.1, while Circuit 13 is expected to saturate at 0 (as explained in Appendix \ref{app:exp_saturation}, the true saturation value here can only be said to be below the systematic statistical finite-sampling bias of 0.0039).
In Appendix \ref{app:exp_saturation} we show numerics of expressibility saturation with respect to two-qubit gate number. We find that different PQCs saturate at different layer numbers and to different expressibility values. This observation may inform the selection of PQCs used in practice.

Consider trying to design a parameterized quantum circuit to maximize expressibility, while maintaining a low depth.
One should choose a circuit that does not saturate at a poor value of expressibility and choose a number of layers that is below that circuit's saturation point. Furthermore, in thinking about using expressibility to choose circuit fragments, saturation might be useful in determining the number of layers used in each circuit fragment.

In the case that a PQC achieves sufficiently favorable expressibility with added layers, it is important to note that
introducing more layers not only increases the depth but also increases the number of circuit parameters.
While having more circuit parameters can increase the dimension of the manifold of states explored,
it may also make the optimization more challenging.
If the states in the smaller manifold (from adding some $k$ layers) are close in distance to the larger manifold (using greater than $k$ layers), then a similar optimization performance can be achieved, while using fewer parameters.
The expressibility value may be an indicator of the representational power of a PQC, informing how many circuit layers are sufficient to use in an application.
For a better understanding of how expressibility (saturation) correlates with the algorithm performance, future investigation is needed.

\bigbreak

\noindent \textbf{{Types of two-qubit operations}}.
Many circuit structures in the benchmark set utilized parameterized two-qubit operations.
These included controlled-Z rotation ($\texttt{CRZ}$) and controlled-X rotation ($\texttt{CRX}$) gates.
Controlled-Z rotation gates tend to feature in many experimental demonstrations as they are natively 
implemented in superconducting architectures.
Several experimental demonstrations have used layers of controlled-Z rotations for variational quantum algorithms \cite{O_Malley2016, otterbach2017unsupervised} .
Controlled-X gates can be constructed by conjugating a controlled-Z rotation by a Hadamard gate.

To compare expressibility and entangling capability of the two types of gates, we considered several pairs of circuit templates that only differ in the type of two-qubit gates used. For example, circuits $5$ and $6$ share a common circuit template, with $5$ using $\texttt{CRZ}$ gates and $6$ using $\texttt{CRX}$ gates in their respective two-qubit entangling blocks. The descriptor values for such circuit pairs are shown in Fig. \ref{table:two_qubit_op_types}. For each pair, circuits employing $\texttt{CRX}$ gates corresponded to more favorable expressibility and entangling capability. 
An explanation might be that $\texttt{CRZ}$ operations in the entangling block commute with each other and thus the effective unitary operation comprised of $\texttt{CRZ}$ gates can be expressed using unique generator terms that are fewer than the number of parameters for these gates. 
Accordingly, the dimension of the manifold of states explored by, say, circuit 5 will be less than the dimension of that explored by circuit 6.
This suggests that, if one is trying to design a PQC to increase expressibility,
it is better to insert single-qubit gates which skew the controlled-gate rotation axis away from the control axis (i.e. the Z-axis).

\begin{table}
\def\arraystretch{1.5}\tabcolsep=8pt
\begin{tabular}{|ccc||ccc|}
% \hline
\multicolumn{3}{c}{\textbf{CRZ}} & \multicolumn{3}{c}{\textbf{CRX}} \\ \hline
\textbf{ID} & \textbf{Expr} & \textbf{Ent} & \textbf{ID} & \textbf{Expr} & \textbf{Ent} \\ \hline
3 & 0.24 & 0.34 & 4 & 0.13 & 0.47 \\ \hline
5 & 0.06 & 0.41 & 6 & 0.004 & 0.78 \\ \hline
7 & 0.10 & 0.33 & 8 & 0.09 & 0.39 \\ \hline
13 & 0.05 & 0.61 & 14 & 0.01 & 0.66 \\ \hline
16 & 0.26 & 0.35 & 17 & 0.14 & 0.45 \\ \hline
18 & 0.24 & 0.44 & 19 & 0.08 & 0.59 \\ \hline
\end{tabular}
\caption{Descriptors computed for six circuit pairs with widths of $n=4$ qubits and depths of $L=1$ layer. Data for each pair are reported in the same row. Each pair assumes the same circuit template with varying two-qubit operations ($\texttt{CRZ}$ or $\texttt{CRX}$ gates).}
\label{table:two_qubit_op_types}
\end{table}

\bigbreak

% Two-qubit operation configuration (v2)
\begin{table}
\def\arraystretch{1.2}\tabcolsep=8pt
\begin{tabular}{|c||cc|}
\hline
\textbf{Configuration} & \textbf{Expr} & \textbf{Ent}  \\ \hline
\text{Nearest-neighbor} & 0.087 & 0.67 \\ \hline
\text{Circuit-block} & 0.015 & 0.80 \\ \hline
\text{All-to-all} & 0.011 & 0.80 \\ \hline
\end{tabular}
\caption{Descriptors computed for circuits (when $n=4$) from Fig. \ref{fig:circuit_configurations} that employ different configurations of two-qubit gates, i.e. nearest-neighbor, circuit-block, or all-to-all.}
\label{table:two_qubit_op_config}
\end{table}

\begin{figure}
\centering
\includegraphics[width=0.48\textwidth]{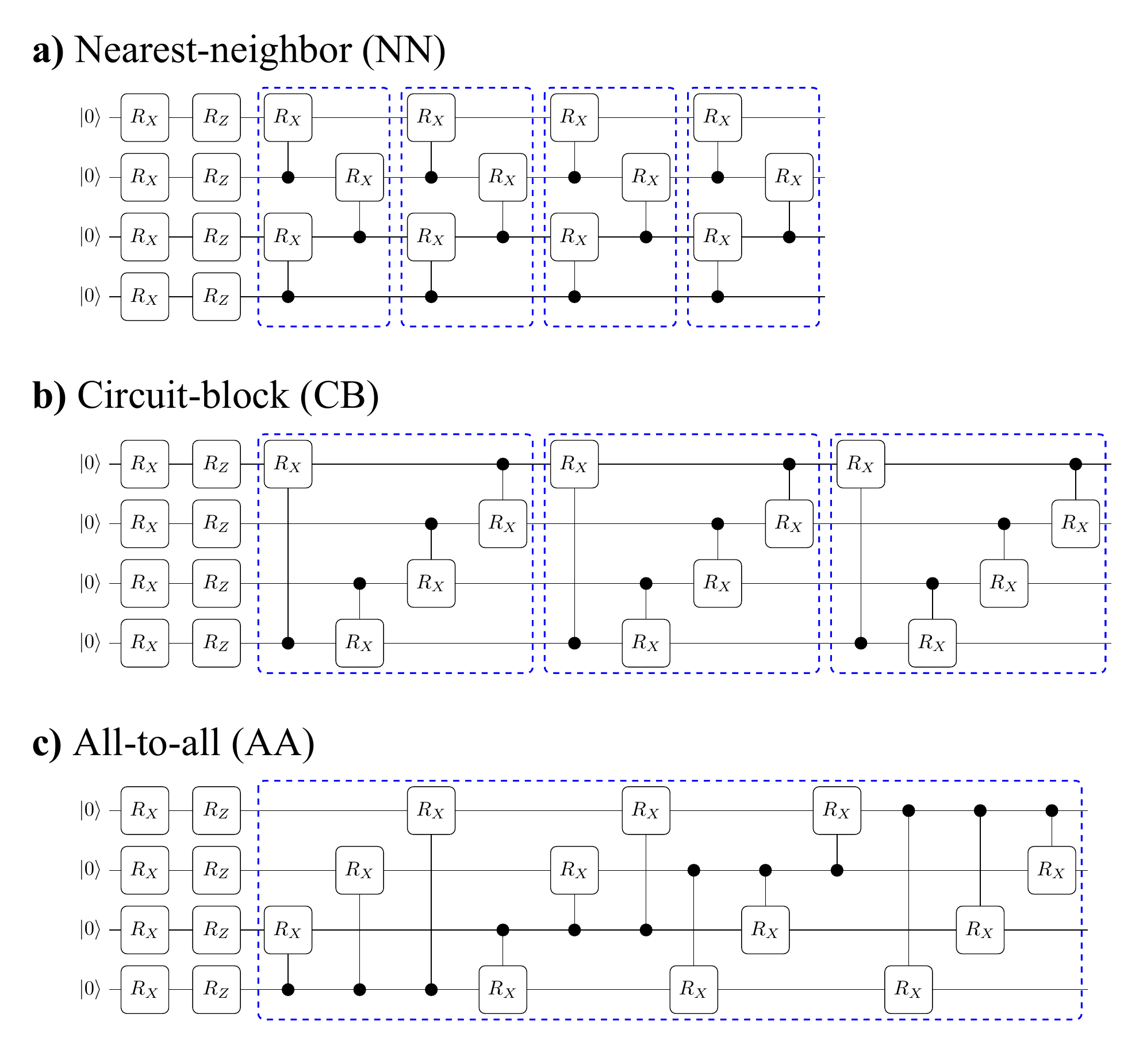}
\caption{Circuits considered for comparing two-qubit interaction configurations: (a) nearest-neighbor, (b) circuit-block, and (c) all-to-all. Two-qubit entangling blocks are shown in blue dashed lines. For a fair comparison, these circuits assume the same number and type of two-qubit operations.}
\label{fig:circuit_configurations}
\end{figure}

\noindent \textbf{{Configurations of two-qubit operations}}.
Though abstract quantum circuits are often depicted in a linear orientation, recent hardware developments allow for more complex qubit topologies, e.g. nearest-neighbor interactions on a two-dimensional lattice or all-to-all qubit interactions, depending on the architecture. To investigate the potential advantages of these qubit topologies, three configurations of two-qubit gates were compared: nearest-neighbor (NN), circuit-block (CB), and all-to-all (AA) configurations. The NN configuration is a natural arrangement of two-qubit operations on a linear array of qubits. The CB configuration is a natural arrangement for an array of qubits that form a closed loop. The AA configuration assumes a fully connected graph arrangement of qubits. To set up a fair comparison, we considered the three circuits shown in Fig. \ref{fig:circuit_configurations} allowing the same single-qubit rotations as well as the same number of two-qubit operations. Qualitatively, the circuit-block configuration from \cite{SchuldCircuitCentric} (in which the range of control is fixed to 1) can be interpreted as an intermediate between the nearest-neighbor and all-to-all configurations. That is, for each circuit block, this circuit structure includes regions of consecutive nearest-neighbor interactions in addition to a non-local interaction to complete the cyclic connectivity.

Both expressibility and entangling capability were computed, as shown in Table \ref{table:two_qubit_op_config}. The AA configuration led to the most favorable expressibility (lowest KL divergence), although the CB configuration had an expressibility value close to that of the AA configuration. Although the NN configuration had the worst expressibility, for the same number of two-qubit operations, it corresponded to the lowest circuit depth. Trends in entangling capability were similar: both CB and AA configurations led to high entangling capability. Therefore, the use of an all-to-all configuration led to both favorable expressibility and entangling capability scores but with a trade-off in the number of parameters, circuit depth, and qubit connectivity requirements. Though slightly less expressible than the all-to-all configuration, the use of the circuit-block architecture led to relatively favorable expressibility and entangling capability, offering a cheaper or more near-term circuit structure alternative.

\bigbreak

% Circuit configurations

% Circuit landscape
\begin{figure}
\centering
\includegraphics[width=0.45\textwidth]{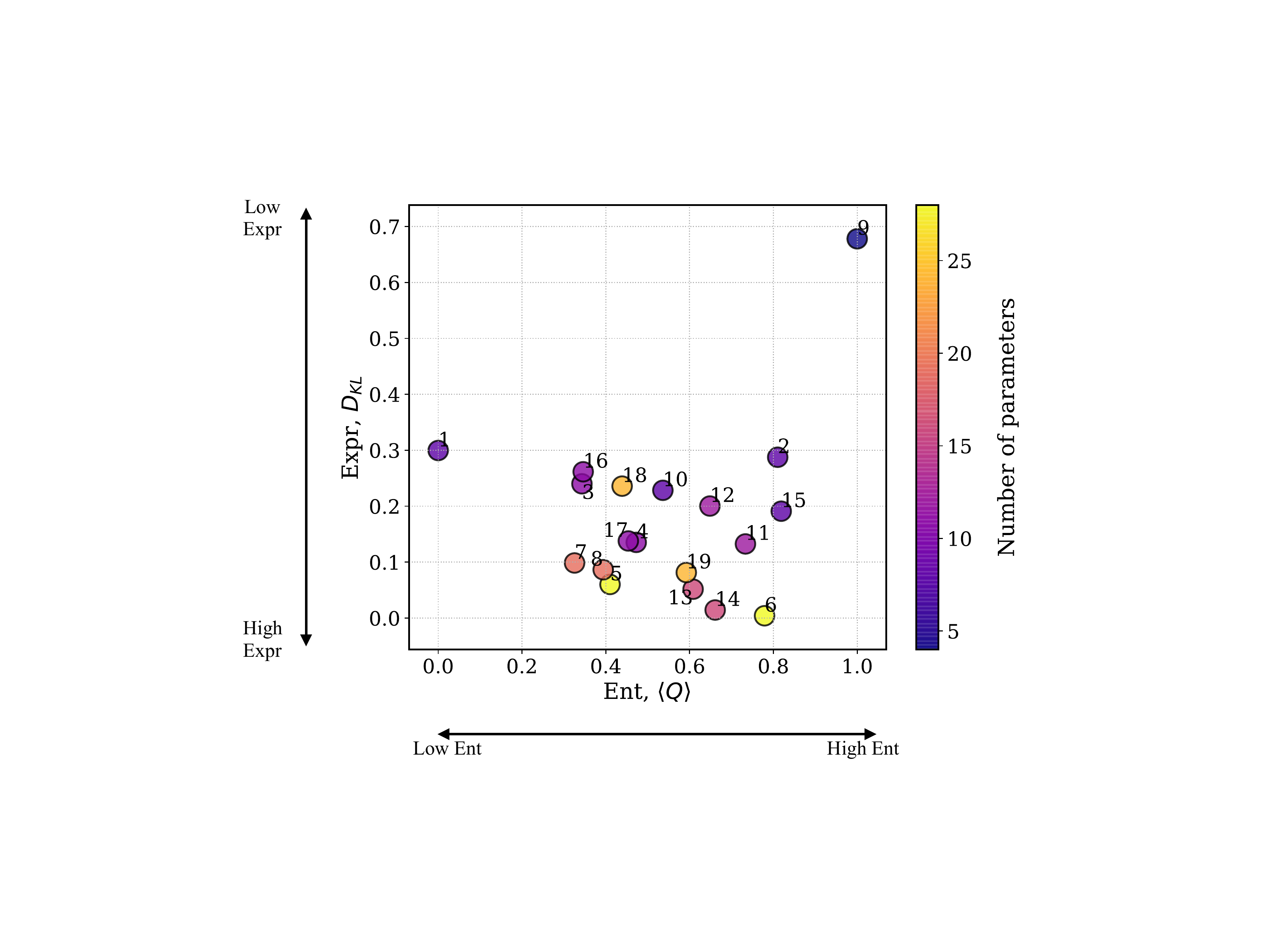}
\caption{Circuit descriptor landscape for circuit instances with width of $n=4$ qubits and depth of $L=1$ layer. Circuits are labeled by their IDs assigned in Fig. \ref{fig:test_circuits}. Marker color indicates the number of parameters associated with the circuit instance.}
\label{fig:exp_ent_map}
\end{figure}

\noindent \textbf{{Multi-descriptor comparison}}. In practice, when selecting a suitable circuit structure for a given application, laying out the ``circuit descriptor landscape'' may be a useful method to identify circuits that have favorable qualities as well as reasonable circuit costs depending on the resource constraints of the quantum hardware. An example of such landscape plot is shown in Fig. \ref{fig:exp_ent_map}. From this plot, for an application in which favorable expressibility can improve the performance, one may consider selecting circuit $6$. However, for a ``cheaper'' alternative in terms of the number of parameters and circuit depth, one may instead choose to use circuit $14$. 

\bigbreak

\noindent \textbf{{Limitations of the method}}. With growing system size, a large number of state samples will be required to estimate each property to a reasonable precision: the expected overlap between states drops exponentially with qubit number. Nevertheless, the theoretical framework from this study can be useful for studying and designing modest-sized circuits that are suitable for NISQ computers. In addition, trends observed in the descriptors for several low-width circuits appear to generalize to larger widths. We also note that estimation of each descriptor may be improved using alternative quantities or numerical methods. For example, the entanglement measure used to quantify the entangling capability is one particular measure of multi-particle entanglement that does not fully characterize the entanglement in the system. Additional entanglement measures \cite{Nielsen2003} can be explored in tandem for an in-depth characterization.

\bigbreak

\section{\label{sec:conclusion}Conclusion and outlook}
In an effort to improve or create HQC algorithms, numerous studies have developed better ansatze, in terms of accuracy, scalability, and implementability on near-term devices for specific applications \cite{Wecker2015, McClean2016, Kandala2017, DallaireDemers2018, Kivlichan2018}. Many proposals have been either theoretically motivated but impractical, or practical but largely \emph{ad hoc}. There is an opportunity and need for developing principled approaches to designing parameterized quantum circuits.

In this work, we presented a theoretical framework to characterize and compare parameterized quantum circuits, independent of the algorithm or application. With one of the descriptors, expressibility, its value was shown to saturate with sufficient depth. We described how the rate of saturation and the saturated value may be useful indicators of the performance of a parameterized quantum circuit and, therefore, may help to design and select such circuits in an application.
In addition, we applied the descriptors to identify useful circuit fragments, in terms of both the gate choice and the configuration of two-qubit operations. Several of these fragments are natural operations for particular quantum hardware (e.g. all-to-all qubit connectivity of ion trap quantum computers) and thus may guide designs of PQCs for experiments on particular devices. However, there still remain open questions and challenges regarding design of PQCs that will be explored in future work.

It remains to understand the connection between expressibility and performance of a PQC in a particular algorithm such as VQE.
This will require a deeper benchmark study that quantifies the correlations between the descriptors and the performance metrics of a given algorithm, e.g. energy errors and/or numbers of function evaluations.
Some degree of expressibility is expected to be necessary in order that a PQC performs well for a variety of Hamiltonians in VQE.
For certain settings, it may be advantageous to exploit symmetry in the problem to design a PQC.
As a simple example, consider a fermionic second-quantized Hamiltonian describing a system with fixed particle number. In this case, the parameterized quantum circuit may be designed to output states only in the proper particle-number subspace, resulting in an unfavorable expressibility. One approach to handling this situation would be to generalize the notion of expressibility to \emph{subspace expressibility}, comparing the distribution of fidelities to that of the Haar distribution over the subspace.  
By understanding the correlations between the descriptors and the performance metrics of a variational HQC algorithm, these descriptors may become useful for guiding the design and compilation of the ansatz. In practice, it may be particularly helpful to derive a unified quantity that combines the descriptors to evaluate a single value that quantifies the capability of a PQC, similar to the ``quantum volume'' \cite{qvolume} for quantifying the capabilities of quantum devices.

Thus far, our descriptors have only been explored for pure-state classical simulations of quantum circuits.
To better understand the performance of parameterized quantum circuits in a more-realistic setting, it may be worth exploring noisy simulations or designing experimental protocols to estimate expressibility (e.g. using SWAP tests to compute state fidelities).

A major challenge in developing and scaling up variational HQC algorithms is the ``barren plateau'' phenomenon highlighted in \cite{McClean2018}. The authors show that the expectation value of the gradient of the objective function rapidly approaches zero with increasing system size when the output states are randomly drawn from an approximate 2-design. 
This shows that expressible circuits must be used with care.
In particular, with an expressible circuit, choosing a random starting point for a VQE optimization is not a reasonable approach.
Methods for circumventing this issue have been proposed \cite{McClean2018,grimsley2018adapt,Grant2019}, suggesting ways to improve the ``optimizability'' of a parameterized quantum circuit.
These techniques indicate a tension between being expressible and being optimizable (i.e. having sufficient variation in the cost function). Both qualities are expected to be important in practice.
This tension points to an opportunity to investigate the proper balance between these two qualities when solving particular problems of interest.

While this study provides only loose design criteria for selecting circuit ansatze, it presents a concrete classical simulation framework for identifying and approaching the said challenges by defining quantities that can be easily computed and compared among various circuit designs or, even, circuit fragments. This can allow for one to evaluate and rank a group of potential circuits based on criteria such as the application-of-interest and/or hardware-of-choice. We intend for this study to be a useful starting point for researchers designing parameterized quantum circuits, as well as for experimentalists benchmarking and developing gates for quantum devices.

%% Acknowledgements 
\begin{acknowledgments}
The authors thank Ryan Sweke for a helpful discussion in connecting ideas in this work to ideas in statistical learning theory. The authors also thank Morten Kjaergaard, Juani Bermejo-Vega, Maria Schuld, and Jonathan Olson for valuable discussions and comments on the manuscript. S. S. also thanks Aram Harrow and Isaac Chuang for helpful discussions on the project regarding frame potentials and parameterized circuit designs respectively during the Quantum Information Science (8.371) course. All the authors thank the colleagues at Zapata Computing for their feedback. A. A.-G. acknowledges support from Anders G. Fr\o seth as well as the Google Focused Award. A. A.-G. acknowledges support from the Vannevar Bush Faculty Fellowship under Award No. ONR 00014-16-1-2008 and the Army Research Office under Award No. W911NF-15-1-0256. S. S. is supported by the Department of Energy Computational Science Graduate Fellowship (DOE CSGF) under grant number DE-FG02-97ER25308.
\end{acknowledgments}

\bibliographystyle{apsrev4-1}
\bibliography{main}

\vspace{1cm}

% Make appendix single-column
\onecolumngrid

\appendix
\section*{Appendix}

\section{Circuit width\label{app:circuit_width_analysis}}

% Width analysis (separated)
\begin{figure}
\centering
\includegraphics[width=0.9\textwidth]{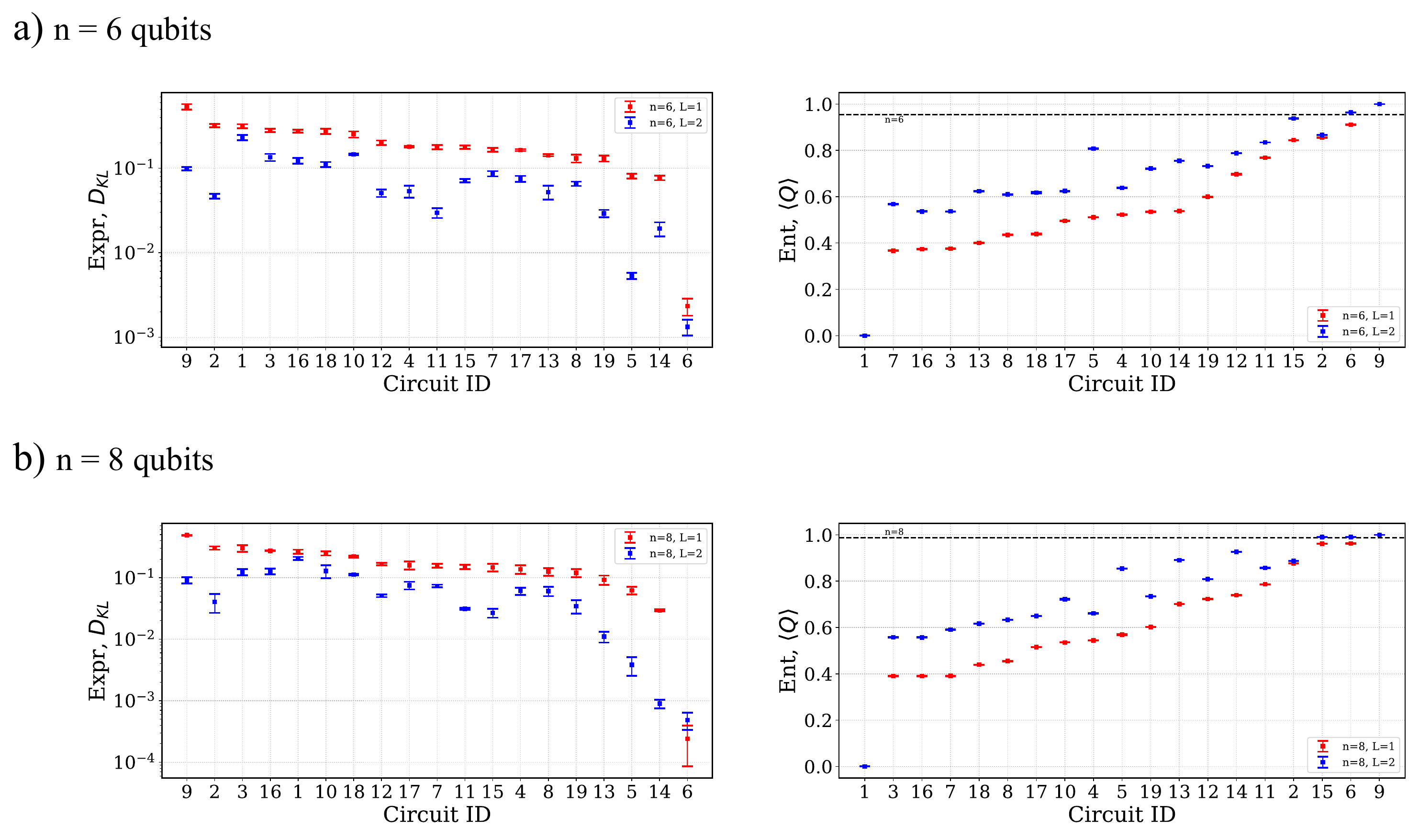}
\caption{Expressibility and entangling capability computed for circuits from Fig. \ref{fig:test_circuits} considering larger circuit widths: (a) $n=6$ and (b) $n=8$ qubits. Marker colors indicate different numbers of circuit layers ($L$) applied, and marker types indicate different circuit widths ($n$). In each plot, circuits are ordered by ascending descriptor values for when $L=1$. Black dashed lines in entangling capability plots for both (a) and (b) show the average $Q$ values for random pure states. Error bars show standard deviations over three independent computations.}
\label{fig:width_analysis}
\end{figure}

This section aims to numerically demonstrate the utility of the theoretical framework for characterizing modest-sized circuits (i.e. with widths and depths suitable for NISQ devices).
We extend the descriptor analysis of the circuit templates by considering their instances at larger qubit numbers: $n=6$ and $n=8$.
For each circuit width, we compute expressibility and entangling capability of the circuit at depths $L=1$ and $L=2$ (in circuit layers).
As observed in Fig. \ref{fig:width_analysis}, the relative (ascending) ordering based on descriptor values as well as trends in the rate of increase for these descriptors with an added circuit layer are largely preserved, up to statistical error.
For example, circuit $9$ is the least expressible for the three different circuit widths considered, and circuit $6$ is the most expressible for all three cases. A similar result is observed for circuits $1$ and $9$ for the entangling capability.
While the raw values of the rates of increase in expressibility or entangling capability change with circuit width, circuit templates at $n=4$ that correspond to large (or small) increases in the descriptor values from $L=1$ to $L=2$ also are templates that correspond to large (or small) increases in the descriptor values for larger qubit numbers.
With circuit templates that exhibit consistent trends in descriptor values over varying qubit numbers, it may be possible to use observations and insights from simulations of smaller circuit instances to infer the performance of the same circuit template with a larger qubit number.

\section{Sample size\label{app:n_samples}}
\begin{figure}
\centering
\includegraphics[width=0.5\textwidth]{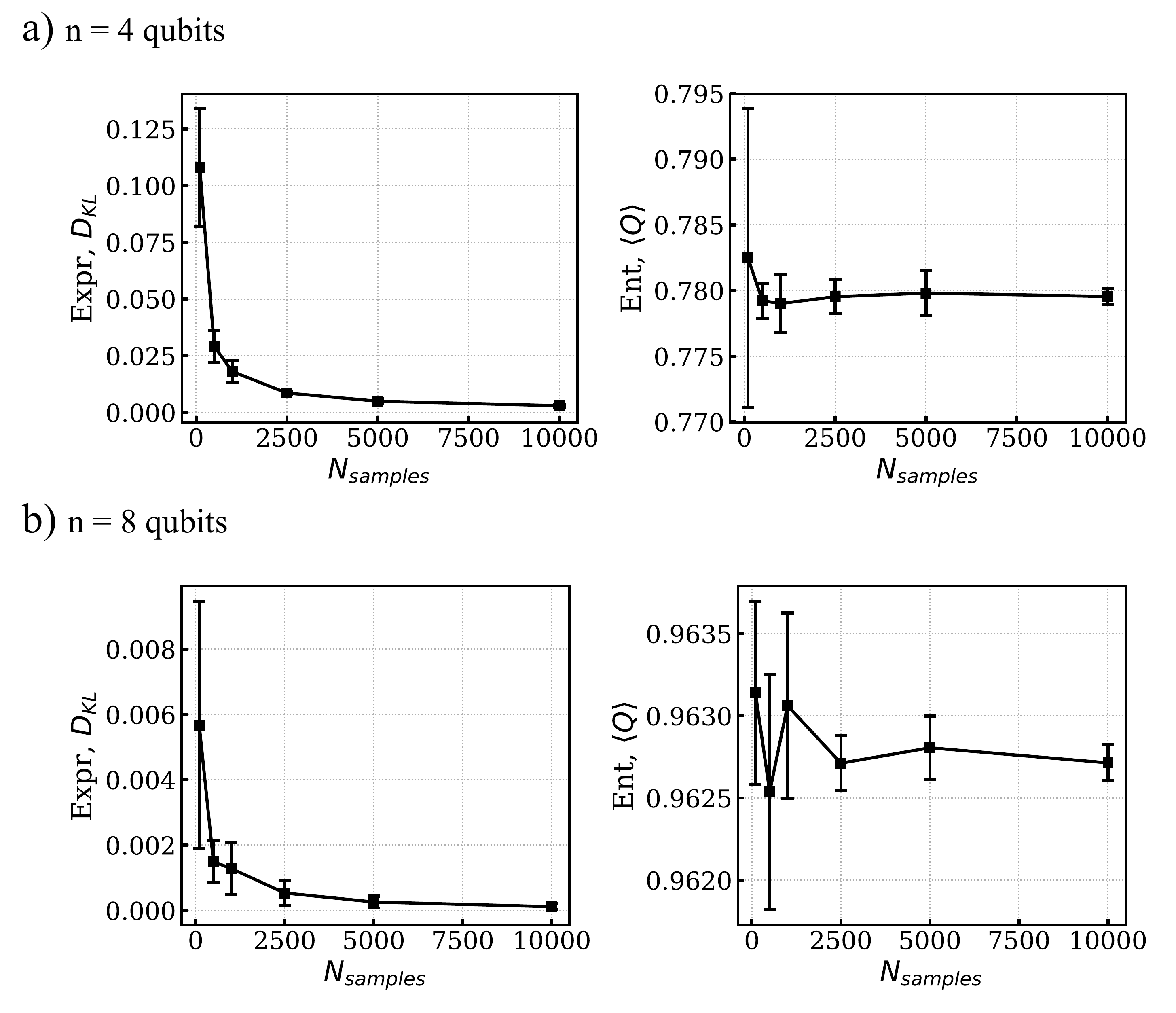}
\caption{Plots showing convergences in both expressibility and entangling capability values for circuit $6$ with increased sample size. Plots (a) and (b) show the descriptor data for $n=4$ and $n=8$ qubits, respectively. Error bars show standard deviations over five independent computations. The term $N_\text{samples}$ refers to the number of sample \emph{pairs} of parameterized states.}
\label{fig:descriptor_convergences}
\end{figure}

The appropriate sample size for estimating expressibility and entangling capability was deduced by applying Chebyshev's inequality. For each circuit simulation in Section \ref{sec:numerical}, $5000$ pairs of states ($10^4$ states total) were sampled to compute the descriptors. In the case of expressibility, this sample size corresponded to estimating the mean of state fidelities within a relative precision of approximately $0.1$, with respect to the variance, with $98 \%$ confidence. Similarly, this sample size corresponded to estimating the average MW measure, or $\langle Q \rangle$, within a relative precision of $0.07$ with $98 \%$ confidence. To empirically justify the sample size, both descriptor values were computed for a single layer of circuit 6, considering two different circuit widths, $n=4$ and $n=8$. Estimated descriptor values are plotted over varying sample size in Figs. \ref{fig:descriptor_convergences}. 
In this figure we observe bias in the estimation of expressibility, in which the magnitude of the bias is pronounced at low sample sizes. Computing an unbiased estimator for entropy (or entropy-based quantities) is a well-known challenge \cite{Chao2003}.
The bias can often be alleviated by a combination of collecting sufficient samples and adding correction terms to the entropy estimate, e.g. Chao-Shen terms for relative abundance and sample coverage \cite{Chao2003}.
We leave such considerations in the context of estimating expressibility to future work.
On the left-hand panels we can observe the bias due to finite sampling of the expressibility value;
the plotted standard deviations do not encompass the subsequent sample means that are estimated using larger sample numbers.

% Sample size analysis figure

\section{Expressibility saturation\label{app:exp_saturation}}

\begin{figure}
\centering
\includegraphics[width=0.9\textwidth]{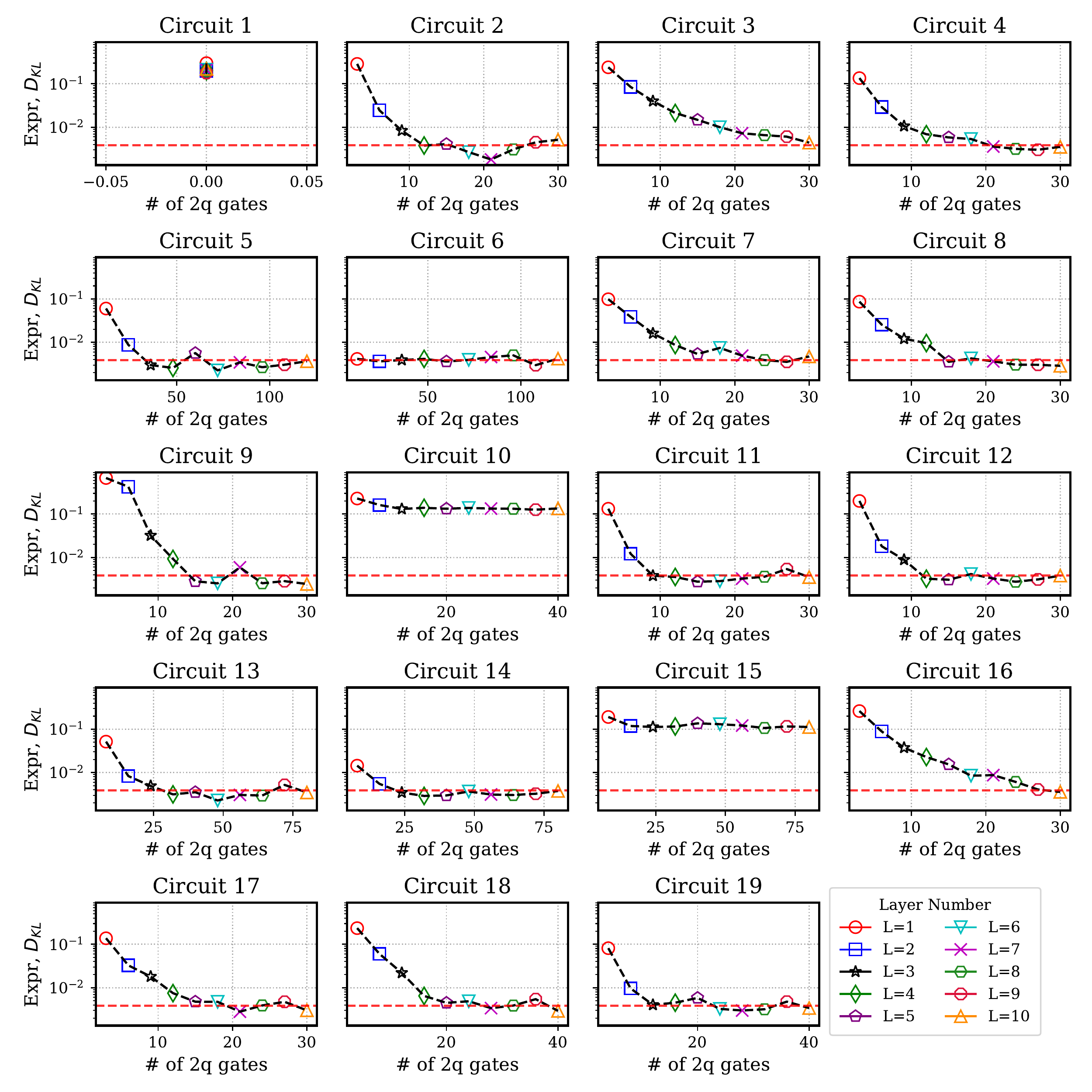}
\caption{Expressibility values plotted against the number of two-qubit gates for each circuit in Fig. \ref{fig:test_circuits} with width of $n=4$ qubits. Markers indicate the layer number. As described in Appendix \ref{app:exp_saturation}, finite sampling introduces a bias in the expressibility estimation. The red dotted line in each subplot shows the estimator's bias introduced in the case of 5000 samples when the true value is zero (i.e. the expressibility of the Haar distribution itself).}
\label{fig:expressibility_saturation_2q_gates}
\end{figure}

In Section \ref{sec:test_circuit_expressibility}, we observed evidence of expressibility values saturating with increased layers. We extend the number of layers $L$ up to 10, to confirm and further investigate this phenomenon. Expressibility values are plotted with respect to the number of two-qubit gates as shown in Fig. \ref{fig:expressibility_saturation_2q_gates}, where the markers indicate the layer number. This allows the reader to visualize the saturating effect while taking into account the total number of two-qubit gates it took to reach a particular expressibility value. The red dotted line in Fig. \ref{fig:expressibility_saturation_2q_gates} indicates the bias in the estimation of the KL divergence due to finite sampling, in which this value is computed by averaging over five sets of $5000$ Haar-random states. This provides a numerical barrier below which we cannot resolve the expressibility values.
In order to better resolve the value of expressibility saturation below this numerical barrier (e.g. in the case of circuit 16), we would either need to generate more samples or use a bias-corrected estimator.

From the plot, we observe different cases of expressibility saturation, varying in the rate of and value at saturation.
For instance, circuits 6, 10, and 15 saturate at nearly the first layer. However, circuits 10 and 15 saturate at unfavorable expressibility values, compared to that of circuit 6. 
Circuit 3 (or, equivalently, 16) does not reach saturation even with ten circuit layers, though the expressibility continues to improve. 
Many of the circuits, regardless of the qubit connectivity or two-qubit gate configuration, reach favorable expressibility values with sufficient depth.
In practice, we may be interested in circuits that correspond to the most favorable expressibility (lowest KL divergence) while maintaining a low number of two-qubit gates. Circuits 11 and 12 saturate at favorable expressibility values using between 10 and 20 nearest-neighbor two-qubit operations. By contrast, circuit 6 reaches a
% low \pj{(as in good or bad?)} % high expr => low KL divergence
favorable expressibility value within a single circuit layer, using 12 two-qubit operations, but several of these gates are non-local, requiring a higher degree of qubit connectivity or a (costly) decomposition into nearest-neighboring gates. This implies that circuits 11 and 12 may be better circuit choices to employ on NISQ devices than circuit 6 is.

% Figure of expressibility saturation

\end{document}